\documentclass[12pt]{article} 
\usepackage[letterpaper, left=1in, right=1in, top=1in, bottom=1in]{geometry} 
\usepackage[utf8]{inputenc}
\usepackage{authblk}
\usepackage{amsmath}
\usepackage{mathrsfs}
\usepackage{mathtools}
\usepackage{bigints}
\usepackage{textgreek}
\usepackage{graphicx} 
\usepackage{subfig}
\graphicspath{{../figures/}} 
\usepackage[font=small,labelfont=bf]{caption}
\usepackage{booktabs} 
\usepackage{makecell} 
\usepackage{multirow} 
\usepackage{xcolor}
\usepackage[colorlinks,allcolors=cyan!70!black]{hyperref} 
\usepackage{setspace}
\usepackage{lineno}
\usepackage[super,numbers,sort,compress]{natbib}
\usepackage{microtype} 

\begin{document}

\title{Microstructure evolution during rapid solidification of hypoeutectic Al-Ag alloys near absolute stability}
\author[a]{Brian Rodgers}
\author[b]{Mingwang Zhong}
\author[a]{Trevor Lyons}
\author[c]{John Roehling}
\author[c]{Joseph T. McKeown}
\author[b]{Alain Karma}
\author[a,d,*]{Amy J. Clarke}

\affil[a]{Colorado School of Mines, Department of Metallurgical and Materials Engineering, Golden, CO, 80401, USA}
\affil[b]{Northeastern University, Department of Physics, Boston, MA, 02115, USA}
\affil[c]{Lawrence Livermore National Laboratory, Materials Science Division, Livermore, CA, 94550, USA}
\affil[d]{Los Alamos National Laboratory, Sigma Manufacturing Science Division, Los Alamos, NM, 87545, USA}

\affil[*]{Corresponding author: amyclarke@mines.edu, aclarke@lanl.gov (A.J. Clarke)}

\date{}
\maketitle
\section*{Abstract}
Microsegregation-free microstructures can form by solidifying at velocities beyond the absolute stability limit ($V_{\text{abs}}$), where solute partitioning is suppressed by a stable, planar solid-liquid interface. Producing such microstructures is of considerable practical interest; however, $V_{\text{abs}}$ typically exceeds the ${\sim}1$~m/s growth rates encountered in additive manufacturing (AM). Here we demonstrate the absolute stability limit can be reached in sufficiently concentrated hypoeutectic Al-Ag alloys at growth rates well below the 1~m/s typically encountered in additive manufacturing. Dynamic Transmission Electron Microscopy (DTEM) of rapid solidification front evolution---following laser spot melting of Al-Ag thin films---combined with postmortem microstructural characterization, enables detailed quantitative comparison with both phase-field (PF) simulations and a sharp-interface linear stability analysis that uses a non-equilibrium, velocity-dependent phase diagram extracted from the PF model. The analysis predicts that $V_{\text{abs}}$ follows a trend similar to that of the miscibility gap, first increasing and then decreasing with Ag concentration. Predicted values of $V_{\text{abs}}$ are in good quantitative agreement with PF simulations over the entire hypoeutectic concentration range and with experiments for three concentrated alloys. These results inform the prediction and control of microstructural development in concentrated alloys near the absolute stability limit under AM conditions. \vspace{12pt}

\noindent \textit{Keywords}: additive manufacturing, rapid solidification, aluminum alloys, in-situ imaging, phase-field model, concentrated alloys

\section{Introduction}

Directional solidification of metal alloys produces microstructural patterns like cells or dendrites when the solid-liquid interface is unstable to perturbations \cite{rappaz2009solidification}. The pattern depends on the solid-liquid interface velocity ($V$), thermal gradient ($G$), and local composition \cite{kurz2023fundamentals}. As the $V/G$ ratio increases, the pattern selected shifts from a planar front to cells, then from cells to dendrites, although the transition from cells to dendrites is somewhat statistical and may not occur immediately when it is favorable \cite{trivedi2003cellular}. 
Cellular interfaces consist of smooth, shallow grooves, and their growth direction is largely controlled by the thermal gradient with only a weak influence of crystalline anisotropy. Dendrites, in contrast, develop deep grooves and secondary side-branches, with the primary growth direction transitioning from the thermal gradient direction at low velocities toward the preferred crystallographic orientation at higher velocities~\cite{akamatsu1997similarity,deschamps2008growth, xing2015phase, tourret2015growth}.
These cellular and dendritic patterns involve solute partitioning during growth, producing a solid with a heterogeneous solute distribution resulting in local chemical and property variations that are generally undesirable in practice \cite{TRIVEDI1987effect}. Suppressing this partitioning requires high interface velocities that
are difficult to achieve, but processes such as additive manufacturing (AM) can produce solidification velocities on the order of 1 m/s or higher, approaching the regime where the planar interface is restabilized, known as absolute stability, which produces a chemically homogeneous solid via partitionless growth. 

Partitioning is reduced when solute trapping occurs as the solid-liquid interface velocity nears the diffusional velocity of solute in the liquid, as described by Aziz et al. \cite{aziz1982model,aziz1988continuous}. The onset of solute trapping is associated with a breakdown of chemical equilibrium at the solid-liquid interface, 
where the system follows a velocity dependent metastable phase diagram rather than the equilibrium phase diagram \cite{perepezko1982use,ji2024microstructure}. In this metastable phase diagram, the solidus and liquidus lines converge toward the $T_0$ line with increasing velocity, progressively reducing partitioning and ultimately stabilizing a planar interface via absolute stability \cite{ji2025phase}.

While absolute stability is achievable during AM, the threshold velocity varies across alloys and compositions thereof and may be unrealistic under typical AM conditions. The initial equations predicting the destabilization of the planar interface at low velocities in dilute alloys were derived by Mullins and Sekerka (MS) \cite{mullins1964stability}. Later analyses by Trivedi et al. extend the determination of planar stability to high velocities with a linear stability analysis 
for dilute alloys yielding the equation: \cite{TRIVEDI1986planar,trivedi1989absolute}  
\begin{equation}
\begin{aligned}
V_\text{abs}=\frac{D_lm_l(V_\text{abs})c_\infty[1-k(V_\text{abs})]}{\Gamma k(V_\text{abs})^2},
\end{aligned}
\label{eq:VabsMullins}
\end{equation}
where $D_l$ is the diffusivity of solute in the liquid, $m_l>0$ is the magnitude of the liquidus slope, $c_\infty$ is the nominal solute concentration, $\Gamma$ is the Gibbs-Thomson coefficient, and $k=c_s^0/c_l^0=c_\infty/c_l^0$ is the partition coefficient, defined as the ratio of the equilibrium solid ($c_s^0$) to liquid ($c_l^0$) concentrations at the interface; both the liquidus slope and partition coefficient are velocity-dependent as a consequence of solute trapping. The term $c_\infty(1-k)/k$ represents the miscibility gap that is equal to the concentration difference $c_l^0 - c_s^0$. Even though Eq. \ref{eq:VabsMullins} is only strictly valid for dilute alloys, its dependence on the miscibility gap suggests that when the miscibility gap narrows significantly, such as in concentrated Al-Ag alloys near the eutectic point, $V_\text{abs}$ may decrease rather than increase with solute content. 

Analysis of interface stability is the basis of most analytical solidification models, such as the Kurz-Giovanali-Trivedi model, which predicts dendrite morphology and converges to the marginal stability model as velocities approach the threshold for planar stability in either direction \cite{kurz1986theory}. This model has since been applied to practical scenarios with high $V$, like welding \cite{rappaz1990analysis} and AM \cite{denonno2024solidification}. However, some alloys do not transition directly to absolute stability as velocity increases. Instead, select compositions experience a region of banding, a solidification pattern wherein the interface alternates between partitioning and partitionless growth \cite{kurz1993AlCu,kurz1995AlCu}. Unlike cells or dendrites, bands form parallel to the solidification front and can rapidly sweep laterally to outcompete other grains during growth \cite{WiezorekAl7CuDTEM,WiezorekAl11CuDTEM,WiezorekAl20CuDTEM}. Whether banding
occurs near absolute stability, and what controls its suppression, depends on the kinetic coefficient and alloy thermodynamics \cite{karma1993interface}, which we examine here by phase-field (PF) simulations with a model recently extended for quantitative predictions of rapid solidification\cite{ji2025phase,mckeown2020imaging,zhong2025quantification,ji2023microstructural,ji2024microstructure}.

The theory of absolute stability has also been extended to account for more complex situations, such as solidification of a multicomponent alloy by averaging the predicted $V_\text{abs}$ for each component \cite{coates1968solid,COLIN2016absoluteternary,tourret2023morphological}. However, the additional considerations of solute trapping and convergence of the solidus and liquidus lines to the $T_0$ line reduce the accuracy of these formulae \cite{tourret2023morphological}. Another limitation is the assumption that the alloy concentration is dilute, which does not apply for many engineering alloys. Even when concentration-dependent parameters are substituted into Eq.~\ref{eq:VabsMullins}, the dilute-alloy framework itself breaks down when the miscibility gap narrows, because the solidus and liquidus slopes are no longer approximately constant. Absolute stability typically requires velocities in excess of 1 m/s, but some alloys have been shown to achieve absolute stability at velocities low enough for most or all of a melt pool to grow as a planar front, including 316L stainless steel, Ni-Nb, and Zn-Ag-Mg \cite{PINOMAA2020absolute316L,chadwick2021development,KARAYAGIZ2020absoluteNiNB,RAMIREZLEDESMA2024absoluteZnAgMg}. However, in none of these systems has the low $V_\text{abs}$ been quantitatively explained through a combination of in-situ observation, quantitative simulation, and theoretical stability analysis.

The Al-Ag system is well suited for such a study. In the hypoeutectic region of its phase diagram, Al-Ag features an almost linear liquidus, but also a strongly curved solidus with an inflection point where the slope decreases dramatically, causing the miscibility gap to narrow significantly at higher Ag concentrations. This thermodynamic feature makes Al-Ag an ideal model system to test the hypothesis that a shrinking miscibility gap drives the reduction in $V_\text{abs}$. While the transition to absolute stability has been observed by in-situ imaging with transparent organic analogues~\cite{LUDWIG1996absolutetransparent}, direct observations in metallic alloys requires higher energy particles than visual spectrum photons~\cite{mckeown2020imaging}. Dynamic Transmission Electron Microscopy (DTEM) of Al-Ag thin films meets this requirement, enabling direct electron imaging of the solid-liquid interface during solidification and measurement of solidification velocities. The thin-film geometry also mitigates the development of convection cells, better enabling comparisons to solidification modeling and analytical predictions. The increased surface area to volume ratio also enhances the effect of the liquid-substrate interface, causing the solid to grow as a meniscus, rather than a flat interface~\cite{CAROLI1986thinfilmMS}.

Here we present three complementary lines of investigation. First, DTEM experiments on Al-Ag thin films with 13.9--23.2~at.\% Ag demonstrate a decrease in $V_\text{abs}$ with increasing Ag content. 
Postmortem (Scanning Transmission Electron Microscopy-Energy Dispersive X-ray Spectroscopy) STEM-EDS characterization provides spatially resolved compositions across the dendrites, serving as quantitative benchmarks for partitioning behavior. Second, quantitative PF simulations reproduce the experimentally observed microstructures and $V_\text{abs}$ using (CALculation of PHase Diagrams) CALPHAD-derived free energies \cite{witusiewicz2004ag,dinsdale1991sgte} as thermodynamic inputs. The simulations reveal a reduced kinetic coefficient ($\mu_k^0 = 0.1$~m/s/K) is required to suppress banding cycles, in qualitative agreement with sharp-interface theory~\cite{karma1993interface,ji2025phase,ji2024microstructure}. Third, we extend the linear stability analysis of the planar interface to concentrated alloys, restricting the analysis to a parameter range where banding is absent. We derive from this analysis an analytical expression for $V_\text{abs}$ valid for an arbitrary non-equilibrium phase diagram with solidus and liquidus slopes that depend both on velocity and concentration, as opposed to just velocity as assumed in Eq. \ref{eq:VabsMullins} for dilute alloys. Combining this analysis with a non-equilibrium phase diagram extracted from the PF model using a recently developed methodology \cite{ji2025phase}, we obtain a prediction of $V_\text{abs}$ as a function of alloy composition that is compared quantitatively to both PF simulations and experiments. Together, these results provide a framework for predicting microstructure selection in concentrated alloys under AM conditions, where dilute approximations are not applicable.

\section{Methods}
\subsection{Dynamic Transmission Electron Microscopy (DTEM)}
The DTEM operates by impinging a laser pulse onto the thin film sample, while simultaneously transmitting an electron beam through the sample to create an image\cite{mckeown2016time,campbell2014DTEM,lagrange2006single,lagrange2008nanosecond,lagrange2015DTEM,mckeown2020imaging,zhong2025quantification,ji2024microstructure}. The laser has a Nd:YAG source with a Gaussian profile and ranges from 2.3-4.9 $\mu$J with a 15 ns pulse duration. The current of an electron beam source operating under typical conditions is insufficient to form an image in the 15 ns of integration time in the DTEM, so extra electrons are excited from the emission tip by another laser \cite{lagrange2008nanosecond}. Similarly, the detector cannot operate fast enough to capture video with 2.5 $\mu$s between different frames and 15 ns integration time. The beam is instead deflected by electrical fields to different regions of the detector to capture the entire movie in one image. This method limits the movies to a selected number of frames (nine in this instance), so experiments are repeated with different offset times for the onset of imaging to capture the full sequence of events \cite{lagrange2015DTEM}. The detector is a CCD camera with 2k by 2k resolution, resulting in a 256 by 256 pixel image for each image in the nine frame movie. 

DTEM samples were prepared by co-sputtering Al and Ag targets onto an amorphous 50 nm thick silicon nitride substrate 0.5 by 0.5 mm wide. The metallic layer created is approximately 100 nm in thickness with nanograins less than 150 $\mu$m in diameter. During laser heating, it is possible volatilization changes the composition, due to the high ratio between surface area to volume. The composition was tested through STEM-EDS scans and no significant variation existed between the heat affected zones and melt pool centers. All sample compositions were hypoeutectic, ranging from 13.9 to 23.2 at.\% Ag.

\begin{figure*}[!ht]
\centering
\captionsetup{singlelinecheck = false, font=footnotesize, labelsep=period}
      \includegraphics[width=0.9\textwidth]{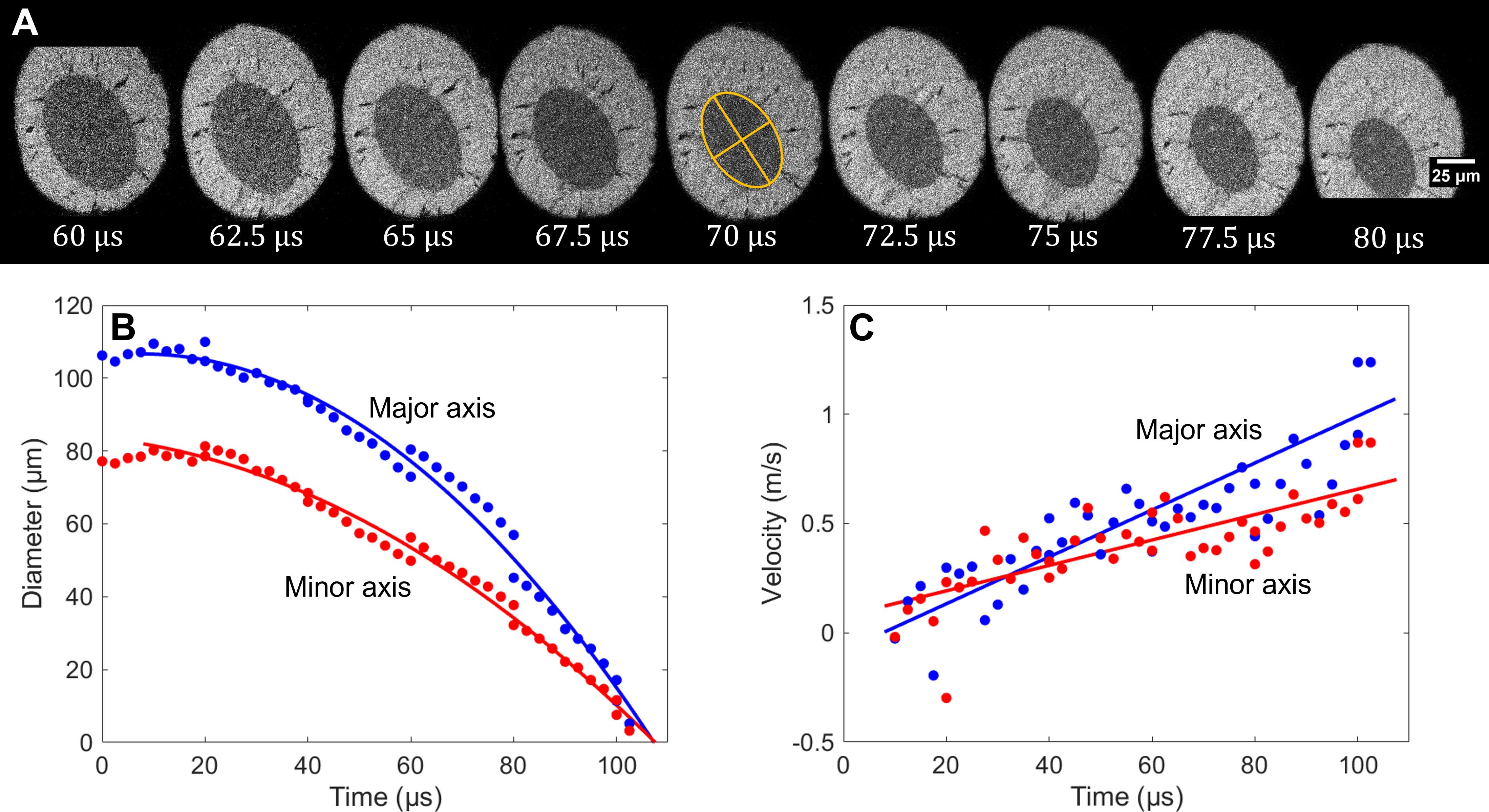}
      \caption{Measurement of interface positions and calculation of solid-liquid interface velocities from DTEM data of Al-13.9 at\% Ag melt pools. (A) Example measurement of major and minor axis lengths during solidification with laser melting of as-sputtered material. (B) Compiled major and minor diameters over time across all DTEM experiments performed in 13.9 at\% Ag melt pools, with a second degree polynomial fit. (C) Line plot of velocities computed by derivation of polynomial fit from (B) to produce lines representing the velocity as a function of time, with a scatter plot of the velocities computed by finite differences included. Velocities are only computed by either method approximately 8 seconds after the laser shuts off to account for the incubation time. The apparent negative velocities in the finite differences plot are a result of scatter, not a retreat of the interface.}
\label{fig:velocity_calcs}
\end{figure*}

The laser beam impinges the DTEM sample surface at a 45 degree angle and produces an elliptical melt pool. The solidification velocity is maximum along the major axis of the ellipse and minimum along the minor axis (Fig. \ref{fig:velocity_calcs}A). Due to the nine frame restriction, solidification velocity is calculated as a single profile by combining the data sets of multiple melt pools (Fig. \ref{fig:velocity_calcs}B). Velocity is computed by fitting a second degree polynomial to the major and minor radii over time and with finite difference methods, similar to prior work \cite{wiezorek2016velocity} (Fig. \ref{fig:velocity_calcs}C). Solidification does not begin immediately when the laser shuts off at $t=0$, but requires an incubation time of approximately 8 $\mu$s.

DTEM melt pools in the Al-Cu system have shown a decrease in velocity with increasing alloy content, as the decreasing melting point leads to larger melt pools for similar laser pulse energies \cite{wiezorek2016velocity,WiezorekAl10CuDTEM,WiezorekAl20CuDTEM}. DTEM melt pools in the Al-Ag system show a similar trend, so velocities are only shown from the lowest Ag content, 13.9 at\% Ag. Melt pools higher in Ag solidify at lower velocities, but begin and end solidification under partitionless planar growth. The actual threshold velocity for partitionless planar growth is therefore not captured by the DTEM velocity measurement in the 16.5 and 23.2 at\% Ag samples, because the transition occurs at a velocity lower than what can be observed. As such, experimental velocity data from the higher Ag content melt pools is not shown so as to prevent confusion. Some DTEM experiments were performed with the same time offset but different laser pulse energies, creating multiple diameter measurements at the same time step. Only the movie with the laser pulse energy closest to the mean was included in these instances, resulting in a range of 4.76-4.89 $\mu$J for the compiled data shown (Fig. \ref{fig:velocity_calcs}).

DTEM melt pools are electron transparent when created, making them ideal for postmortem scanning and conventional transmission electron microscopy, (S)TEM, characterization. Complementary microstructural characterization performed in this work includes TEM, STEM, and EDS performed with an FEI Talos F200X operating at 200 kV or a Titan 300 operating at 300 kV.


\subsection{Phase-Field (PF) Modeling}
In this study, we employed a quantitative PF model to simulate rapid solidification processes in concentrated binary alloys \cite{ji2023microstructural,ji2024microstructure,ji2025phase}, characterized by far-from-equilibrium conditions at the solid-liquid interface with solidification velocities ranging from $\mu$m/s to m/s. Our model leverages a variational framework, introducing nonlinear diffusivity across the solid-liquid interface, thus circumventing the need for the anti-trapping current commonly applied in other models \cite{karma2001phase,echebarria2004quantitative}. This PF model represents solidification using a scalar phase-field variable, $\phi$, assigning values of +1 for the solid phase and -1 for the liquid phase. For concentrated alloys, Gibbs free energies of solid and liquid phases, denoted as $G_s(c,T)$ and $G_l(c,T)$ respectively, were obtained from CALPHAD (thermodynamic) databases \cite{witusiewicz2004ag,dinsdale1991sgte} and expressed as functions of Ag composition $c$ and temperature $T$. The free-energy density interpolates between these two phases as:

\begin{equation}
\begin{aligned}
F_{AB}(\phi, c, T) = \frac{1 + g(\phi)}{2} f_s(c,T) + \frac{1 - g(\phi)}{2} f_l(c,T) + f(\phi),
\end{aligned}
\label{eq:fab}
\end{equation}
where $f_s=G_s/v_0$ and $f_l=G_l/v_0$ represent the free-energy densities with $v_0$ denoting the molar volume, $f(\phi)=h(-\phi^2/2+\phi^4/4)$ is the double-well potential ensuring bistable solid and liquid states, and $g(\phi) = 15(\phi-2\phi^3/3+\phi^5/5)/8$ is the interpolation function. This choice of $g(\phi)$ guarantees $g'(\pm1)=g''(\pm1)=0$ and $g(\pm 1)=\pm1$, allowing $F_{AB}$ to achieve local minima at the solid ($\phi=1$) and liquid ($\phi=-1$) states. Specifically, the energy barrier height $h$ of the double-well potential is given by $h=\gamma_0 a_1^0 W$, where $\gamma_0=\Gamma L/T_M$ is the solid-liquid interface free energy, $\Gamma$ is the Gibbs-Thomson coefficient, $L$ the latent heat of fusion, $T_M$ the melting point of pure Al, and $a_1^0=2\sqrt{2}/3$ is a constant. The interface thickness is defined as $W=S W_0$, where $S$ adjusts its magnitude and $W_0=1$ nm represents the atomistic interface width.

With the free-energy density established, the PF evolution equation is expressed as: 
\begin{equation}
\begin{aligned}
\tau(\vec{n}) \frac{\partial \phi}{\partial t} =& \vec{\nabla} \cdot\left[W(\vec{n})^{2} \vec{\nabla} \phi\right] + \sum_{i}^{x,y,z}\left[\partial_{i}\left(|\vec{\nabla} \phi|^{2} W(\vec{n}) \frac{\partial W(\vec{n})}{\partial\left(\partial_{i} \phi\right)}\right)\right] \\
&+ \phi - \phi^{3} +\frac{1}{2h}g'(\phi)\left[ f_l(c,T) - f_s(c,T) \right],\\ 
\end{aligned}
\label{eq:phi}
\end{equation}
where the first two terms represent the gradient energy at the solid-liquid interface. Anisotropy in the interfacial free energy $\gamma(\vec{n})=\gamma_0a_s(\vec{n})$ and kinetic coefficient $\mu_k(\vec{n})=\mu_k^0a_k(\vec{n})$ is modeled by selecting the time constant $\tau(\vec{n})=a^2_s(\vec{n})/a_k(\vec{n})$ and interface width $W(\vec{n})=W a_s(\vec{n})$, where $a_s(\vec{n})=1-3\epsilon_s+4\epsilon_s(n_x^4+n_y^4+n_z^4)$ and $a_k(\vec{n})=1-3\epsilon_k+4\epsilon_k(n_x^4+n_y^4+n_z^4)$, with $\vec{n}=(n_x,n_y,n_z)$ as the unit normal vector to the interface.

The concentration field $c$ evolves according to: 
\begin{equation}
\begin{aligned}
\frac{\partial c}{\partial t} = \vec{\nabla} \cdot \left[ \frac{D_l^0q(\phi) c (1 - c)v_0}{RT_M} \vec{\nabla} \left( \frac{1+g(\phi)}{2}\partial_cf_s(c,T) + \frac{1-g(\phi)}{2}\partial_cf_l(c,T) \right) \right],
\end{aligned}
\label{eq:c}
\end{equation}
where $R$ is the ideal gas constant, $D_l$ the diffusion coefficient, and $q(\phi)=A(1-\phi)/2 -(A-1)(1-\phi)^2/4$ (with $A=1,6,12$ for $S=1,3,5$ respectively) interpolates diffusivity between solid and liquid phases, ensuring it reduces to Fickian diffusion in the liquid.

Our PF simulations are implemented in C++ and executed on massively parallel graphical processing units (GPUs), utilizing the Compute Unified Device Architecture (CUDA) framework. Table \ref{tab:parameter} lists all simulation parameters. For simplicity, the thermal diffusivity was taken as uniform, averaging solid ($\sim 6\times10^{-5}$ m$^2$/s) and liquid ($\sim 3.5\times10^{-5}$ m$^2$/s) values \cite{leitner2017thermophysical}. Additionally, the Gibbs-Thomson coefficient $\Gamma$ was slightly adjusted to achieve optimal alignment with our experimental observations.

\begin{table}[!ht]
\captionsetup{justification=centering}
\caption{\textbf{Materials and simulation parameters}}
\makebox[\textwidth][c]{
\begin{tabular}{{cll}} 

\toprule
{Symbol} & {Description} & {Value} \\ 
\midrule
{$D_l^0$} & Diffusion coefficient of Ag in the liquid & $5\times 10^{-9}$ m$^2$/s \cite{engelhardt2016AlAgliqdiffusion} \\
{$G$} & Temperature gradient & $5\times 10^{6}$ K/m \cite{pinomaa2020phase} \\
{$L$} & Latent heat of fusion & 10470 J/mol \cite{karma1993interface} \\
{$c_p$} & Heat capacity & 30 J/mol K \cite{karma1993interface} \\
{$D_T$} & Thermal diffusivity & $4.7\times 10^{-5}$ m$^2$/s \cite{leitner2017thermophysical} \\
{$\Gamma$} & Gibbs-Thomson coefficient & $1.3\times 10^{-7}$ K$\cdot$m \cite{kurz1992fundamentals} \\
{$\mu_k^0$} & Interface kinetic coefficient & 0.1 or 0.5\cite{ji2025phase} m/(s$\cdot$K) \\
{$W$} & Interface width & 1 nm ($S=1$) \\
{$\epsilon_k$} & \begin{tabular}{@{}c@{}}Kinetic anisotropy strength \end{tabular} & 0.1 \cite{ji2025phase}\\
{$\epsilon_s$} & \begin{tabular}{@{}c@{}}Interface free-energy anisotropy strength \end{tabular} & 0.012 \cite{ji2025phase} \\
{$\Delta x$} & Grid spacing & 0.8 $W$\\
{$\Delta t$} & Time step of evolution & 0.096 $W^2/(\Gamma\mu_k^0)$\\
\bottomrule
\end{tabular}}

\label{tab:parameter}
\end{table}

\section{Results and Discussion}

\subsection{Experimental observations}
Experimental and theoretical results show there is a decrease in partitioning as velocity increases, i.e. the partitioning coefficient approaches unity \cite{kittl2000complete}. Marginal interface stability (MS) analysis predicts a threshold velocity of zero for absolute stability if there is no partitioning, due to the elimination of constitutional undercooling. Therefore, abnormal reductions in the degree of partitioning related to the unusual shape of the phase diagram for Al-Ag could be an explanation for the low threshold velocity for absolute stability. 

\begin{figure*}[!ht]
\centering 
    \captionsetup{singlelinecheck = false, font=footnotesize, labelsep=period}
    \includegraphics[width=1.0\textwidth]{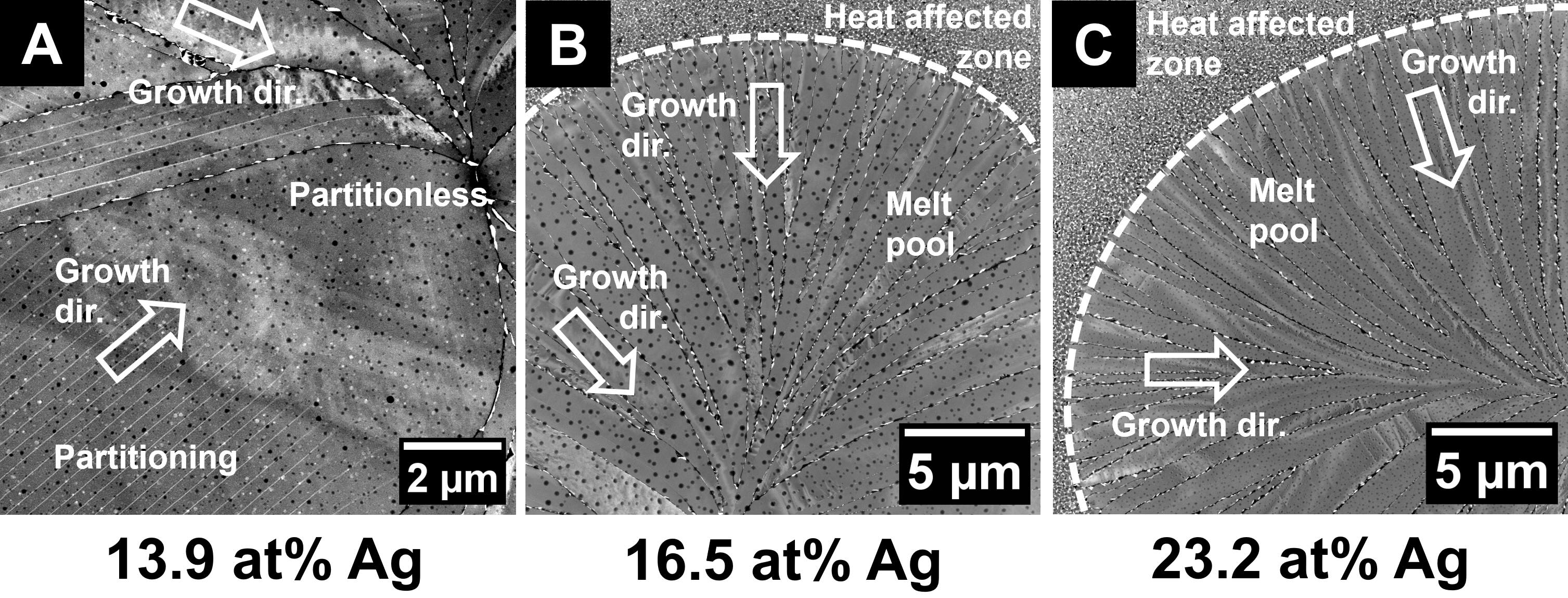}
    \caption{STEM-HAADF (High Angle Annular Dark Field) images of solidification patterns formed during rapid solidification in DTEM samples of Al-Ag alloys with varying Ag solute concentrations, showing a decrease in the tendency towards partitioning growth with increasing solute concentration. All three concentrations form precipitates in the solid state along the grain boundaries during cooling after solidification. Pores also form in all three compositions, appearing as dark or black circle in the images. These pores form due to outgassing from the substrate, and are physical rather than microstructural features. (A) The low concentration, 13.9 at.\% Ag, begins solidification as a dendritic array (i.e., partitioning growth) and transitions to cellular, then planar growth near the end of solidification (i.e., partitionless growth). (B) The intermediate composition, 16.5 at.\% Ag, is fully planar throughout solidification. Note the porosity generated during the DTEM experiment by laser melting associated with sample fabrication, which appears as dark circles, and the precipitates that form in the solid state. (C) The high composition, 23.2 at.\% Ag, is also fully planar.}
    \label{fig:pattern_summary}
\end{figure*}

\begin{figure*}[!ht]
\centering
    \captionsetup{singlelinecheck = false, font=footnotesize, labelsep=period}
    \includegraphics[width=0.9\textwidth]{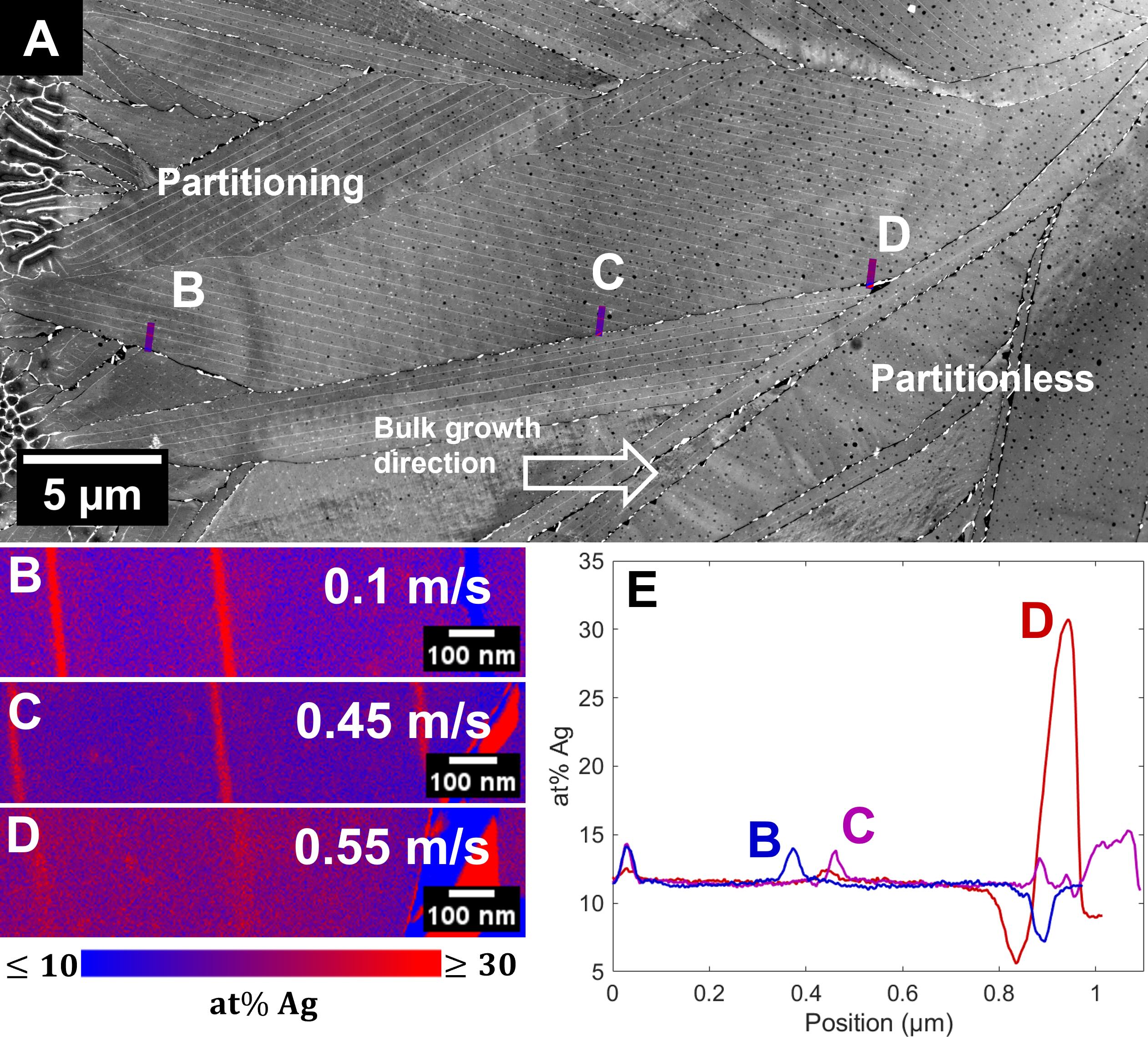}
    \caption{Partitioning behavior of dendrites as velocity increases from left to right of the overview image for the 13.9 at.\% Ag alloy. (A) STEM-HAADF image overview of dendritic to planar solidification as velocity increases. Locations corresponding to composition scans are overlaid. (B,C,D) STEM-EDS maps of marked locations in (A) showing elemental distributions, with blue and red representing Al and Ag, respectively. (E) Quantitative representation of Ag partitioning between the dendrite cores and interdendritic regions, decreasing as velocity increases. The reference frame for the graphed compositions is from left to right in the EDS maps, such that the left edge of the mapped regions corresponds to zero on the graph. The large peaks and valleys correspond to $\gamma$ precipitation along the grain boundaries, while the small peaks correspond to interdendritic regions.}
    \label{fig:dendies}
\end{figure*}

The solid-liquid interface velocity, i.e. solidification velocity, is greatest for the lowest Ag composition, 13.9 at\% with a maximum velocity of approximately 1.2 m/s, as shown in Fig. \ref{fig:velocity_calcs}. This is also the only composition to display partitioning, meaning some solidification occurs below the absolute stability threshold. The majority of the melt pool is a dendritic region, Fig. \ref{fig:pattern_summary}A and \ref{fig:dendies}A, although secondary dendrite arms do not form due to the high solidification velocity \cite{lavernia2010rapidsecondarysuppressed}. Despite lacking secondary dendrite arms, this pattern is still identifiable as dendritic by the misalignment with the thermal gradient and relatively constant primary arm spacing, 300-450 nm, within a grain. Lower interface velocities are measured for the Al-16.5 at\% Ag and Al-23.3 at\% Ag sample conditions, but have passed the absolute velocity threshold, as evidenced by their lack of partitioning, indicating a decreased threshold velocity for absolute stability. The decrease in threshold velocity with increasing composition may be unexpected when examined with the commonly used MS analysis, where $V_\text{abs}$ increases with composition if all other parameters are constant. When diffusivity depends on concentration in the MS analysis, as in Engelhardt et. al. \cite{engelhardt2016AlAgliqdiffusion}, the predicted $V_\text{abs}(C)$ produces a peak at 13.6 at\% Ag and then decays. Despite capturing the correct trend, the predicted absolute velocities overestimate experimental measurements. These discrepancies can be attributed to the unusual shape of the Al-Ag phase diagram, where the liquidus and especially the solidus change slope with composition, causing the miscibility gap to narrow. While the decrease in the threshold velocity with increasing composition may be unexpected at first glance, the sequence of patterns is similar to the expected order for an accelerating interface in non-banding alloys. 


For the low composition (13.9 at\% Ag), the interface is immediately unstable and develops into a dendritic array. The center of the melt pool solidifies partitionlessly, indicating planar growth past the absolute stability threshold. The transition to absolute stability takes place at a velocity of 0.5-0.7 m/s. A few microns of transitory cellular growth between the dendritic and planar regions can be seen with a change in orientation of the partitioning phases (Fig. \ref{fig:pattern_summary}A). This sequence of growth morphologies is expected for an accelerating interface in an alloy that does not form bands. Both of the higher Ag concentrations, 16.5 and 23.2 at\% Ag, begin and end solidification as a partitionless planar front in the absolute stability regime (Fig. \ref{fig:pattern_summary}B,\ref{fig:pattern_summary}C). The lack of a similar transition region makes it impossible to determine the exact threshold velocity for absolute stability with these experiments, but does show the value is very low, less than 0.1 m/s.

Two microstructural features, precipitates and porosity, are present in the micrographs in Fig. \ref{fig:pattern_summary} and are not considered when describing the solidification behavior. Absolute stability initially forms a supersaturated solid, which provides the driving force for the formation of the precipitate, $\gamma$, with grain boundaries serving as nucleation sites for the solid state reaction. The equilibrium secondary phase, $\text{AlAg}_2$ or $\gamma$, has a hexagonally close packed (HCP) structure with an atomic density differing by 1.1 \% from the face center cubic (FCC) matrix phase. A metastable $\gamma'$ phase may also form. The two precipitates are both HCP and differentiated by the structure of the respective precipitate/matrix interfaces. The distinction between the two is not necessary to analyze the solidification behavior, so all precipitates are referred to here as $\gamma$ for clarity and convenience. Separately, porosity appears as a spherical feature in the solid matrix. These pores form through the release of trapped gases from the substrate through the liquid layer when it is still hot, and are purely physical, not chemical features.

Experimental determination of the partitioning coefficient is limited to the one composition, 13.9 at\% Ag, where observed solute partitioning exists. The partitioning is measured on the order of the primary dendrite arm spacing, 300 to 450 nm, with the velocity dependent coefficient measured for three different velocities. This is done with the ratio between the composition of the interdendritic peaks and the base dendritic phase in the STEM-EDS maps, Fig. \ref{fig:dendies}. In reality, the solid-liquid interface does not proceed as a flat wall, but instead has curvature due to the interface with the substrate \cite{CAROLI1986thinfilmMS}. This curvature may cause the observed compositions between the dendrite cores and interdendritic regions to differ from the semi-infinite, i.e. 3D, case as a result of geometric effects. Only the peaks in the first 0 to 0.7 microns of the composition scans are used to determine the partitioning coefficient. Data greater than 0.7 are omitted, because they capture the composition of the intergranular region, which undergoes solid state diffusion. Consistent with expected behavior, the experimentally measured partitioning coefficient approaches unity as the interface velocity increases. At a velocity of 0.1 m/s, near the onset of solidification, the partitioning coefficient is 0.82, increasing to 0.92 at a velocity of 0.55 m/s. This reduction in partitioning can be seen directly by the composition maps and the quantitative line-outs in Fig. \ref{fig:dendies}. Compared to the equilibrium partitioning coefficient of 0.48, partitioning coefficients as high as 0.82-0.92 represent a marked decrease in the degree of microsegregation during solidification.

Velocity dependent thresholds for absolute stability underestimate $V_\text{abs}$ calculated via equation \ref{eq:VabsMullins} for 13.9 at\% Ag when using a partitioning coefficient of 0.92. Therefore, using experimentally measured, non-equilibrium partition coefficients in the MS model does not produce accurate predictions for $V_\text{abs}$ in the Al-Ag system. The MS model was derived under the assumption that the alloy is dilute and the solidus and liquidus slopes are relatively constant. An alloy with 13.9 at\% Ag cannot be considered dilute, and the Al-Ag system has significant curvature in the liquidus and especially solidus lines. These experimental results show the necessity of deriving a more accurate formula for analytical predictions of $V_\text{abs}$ for concentrated alloys. 

\subsubsection{Reheating effects}

In practice during AM, multipass welding, or heat treatment of either, as-solidified material will undergo additional heating in the solid state, promoting further microstructural evolution. Solidifying with absolute stability creates a supersaturated solid solution that readily forms precipitates. For example, all compositions studied here developed $\gamma$ at the grain boundaries. Some DTEM melt pools were made to overlap prior melt pools, and the melt pools reheated this way display unique microstructures. In 16.5 at\% Ag, $V_\text{abs}$ is low enough to entirely lack solidification partitioning. However when the interface of a melt pool crosses into the area of a pre-existing melt pool, the interface is briefly unstable, seen as  1-2 microns of cellular morphology. This brief destabilization does not occur when remelting an as-solidified region at 23.2 at\% Ag, where $V_\text{abs}$ is lower. Therefore, the effect of remelting as-solidified, rather than as-sputtered material, is directly related to interface stability. 

\begin{figure*}[!ht]
\centering 
    \captionsetup{singlelinecheck = false, font=footnotesize, labelsep=period}
    \includegraphics[width=0.9\textwidth]{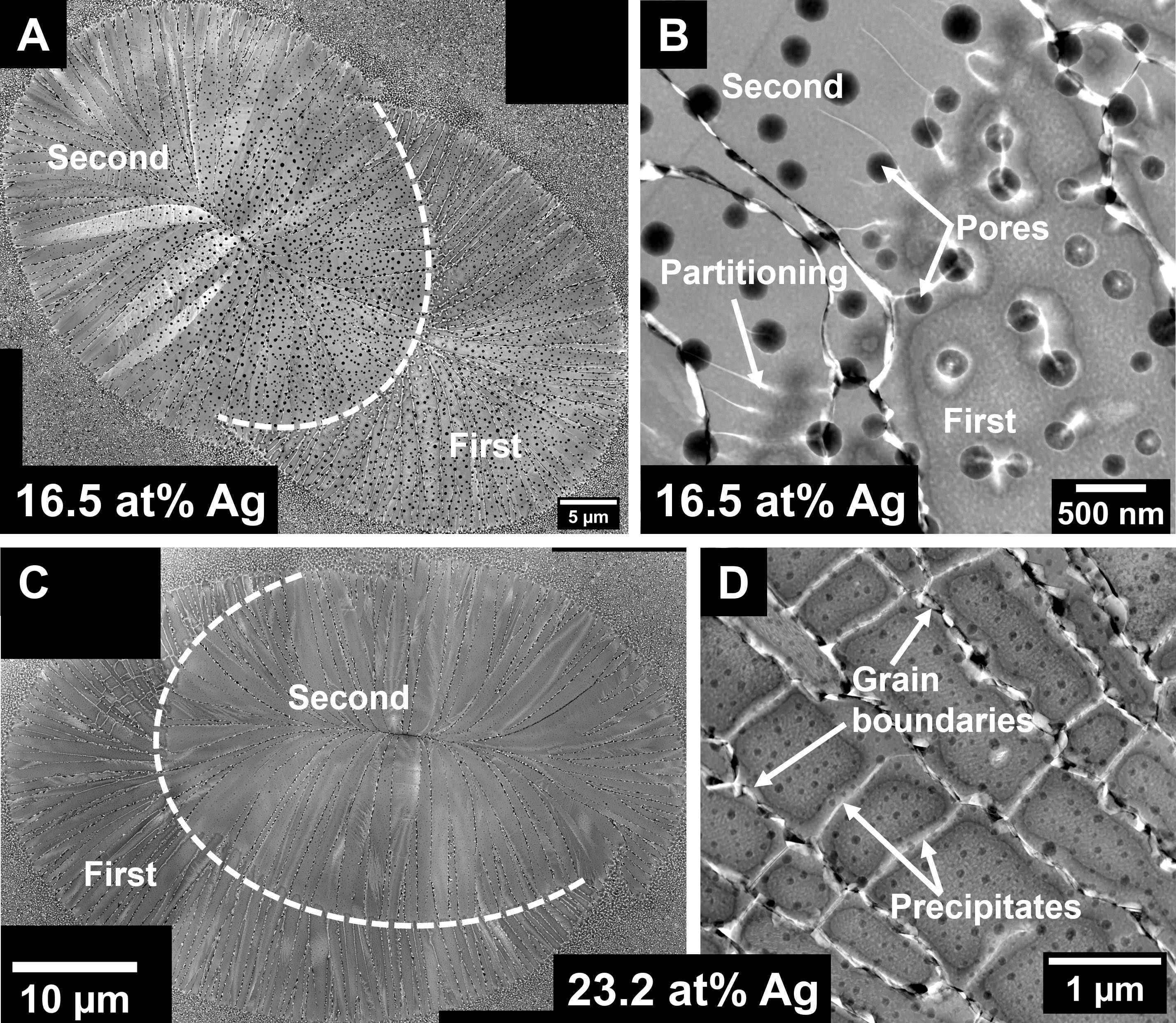}
    \caption{STEM-HAADF images summarizing the effects of remelting on the more concentrated Al-Ag alloys wherein the interface is fully stable throughout solidification when melting as-sputtered material. (A) A stitched image showing the overlapping melt pools in 16.5 at\% Ag and the order in which they were formed. (B) A higher magnification image of the interface at the overlapping region showing the brief destabilization of the interface allowing for partitioning. (C) A stitched image showing overlapping melt pools in 23.2 at\% Ag, which do not experience the brief interface destabilization overlapping melt pools in 16.5 at\% Ag display. (D) A higher magnification image highlighting the unique precipitate structure formed after reheating the melt pool in 23.2 at\% Ag. More precipitation occurs in this higher composition because the supersaturation created by solidifying partitionlessly under absolute stability is greater.}
    \label{fig:secondary_heating}
\end{figure*}

The exact nature of the effect temporarily destabilizing the interface at remelted areas is unknown. Conversely, the increased Ag concentration in 23.2 at\% Ag leading to increased solid state supersaturation promotes further precipitation. While all compositions develop $\gamma$ along the grain boundaries while cooling after solidification, there is limited precipitation within the grain interiors. With the increased supersaturation, the interiors of the grain boundaries develop precipitates as well. Precipitation at the grain interiors does not occur homogeneously; dislocations act as the nucleation sites. Dislocations are present within the grain interiors at all compositions, but are only precipitated upon for the highest alloy composition studied here. The precipitation behavior and solid state phenomena in these alloys under AM conditions in DTEM will be expanded upon in a future paper, including the brick-like structure formed in Fig. \ref{fig:secondary_heating}D. 

\begin{figure*}[!hb]
\centering
      \captionsetup{singlelinecheck = false, font=footnotesize, labelsep=period}
      \includegraphics[width=0.9\textwidth]{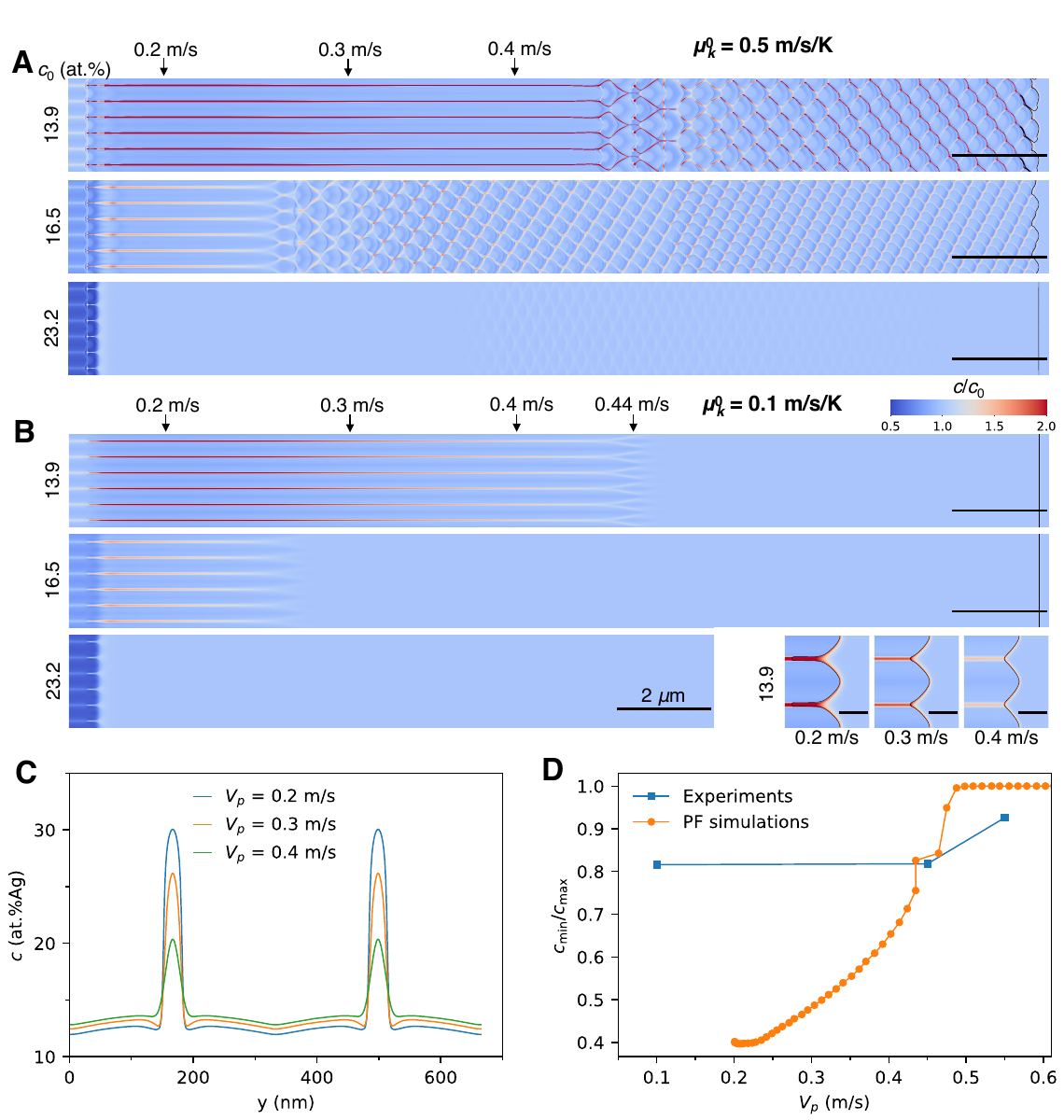}
      \caption{PF simulations of concentrated Al-Ag alloys. (A) Growth morphologies at various nominal Ag concentrations ($c_\infty$) for the kinetic coefficient $\mu_k^0=$ 0.5 m/s/K. The pulling velocity is obtained from experimental measurement (blue line in Fig. \ref{fig:velocity_calcs}C) (B) Growth morphologies at various Ag concentrations for $\mu_k^0=$ 0.1 m/s/K. The three insets show solidification front morphologies at various velocities for 13.9 at.\%Ag, with the black contour denoting the solid-liquid interface. Scale bars: 200 nm. (C) Ag concentration profile across the dendrites for the simulation in (B) at 13.9at.\% Ag. (D) Ratio of the minimum to maximum solute concentration, $c_\text{min}/c_\text{max}$, versus pulling velocity $V_p$ for 13.9at.\% Ag. The experimental measurements are obtained from Fig. \ref{fig:dendies}.}
\label{fig:microstructures}
\end{figure*}

\subsection{Computational and theoretical analysis}
\subsubsection{PF simulations}
To gain deeper insights into the microstructural evolutions observed experimentally, we employed a PF model designed specifically for the rapid solidification of binary alloys. The details of this model are described in the Methods section, with further elaboration provided by Ji et al. \cite{ji2023microstructural,ji2024microstructure, ji2025phase}. Using experimentally measured solidification velocities along the major axis of the melt pool (Fig. \ref{fig:velocity_calcs}C) and assuming a constant thermal gradient of 5 K/$\mu$m, our simulation results show dendritic structures form at low velocities, as illustrated in Fig. \ref{fig:microstructures}. With increasing velocities, the dendrites become oscillatory at a typical kinetic coefficient ($\mu_k^0$) of 0.5 m/s/K for both 13.9 at.\% Ag and 16.5 at.\% Ag, eventually transitioning into banded structures (Fig. \ref{fig:microstructures}A). Notably, the velocity at which dendrites become unstable decreases from 0.41 m/s (13.9 at.\% Ag) to 0.27 m/s (16.5 at.\% Ag). For the highest Ag concentration studied, the interface remains planar at low velocities, but becomes unstable with slight partitioning as velocity increases. While our simulations did not explore extremely high velocities (e.g., 10 m/s), it is anticipated that banded structures will disappear at some critical higher velocities.

Reducing the kinetic coefficient $\mu_k^0$ to 0.1 m/s/K (Fig. \ref{fig:microstructures}B) yields microstructures in closer agreement with experimental observations. For 13.9 at.\% Ag, dendritic structures transition directly to a planar front at approximately 0.44 m/s. As the Ag concentration increases, the absolute stability velocity ($V_\text{abs}$) decreases markedly—dropping to approximately 0.27 m/s for 16.5 at.\% Ag and becoming nearly indiscernible for 23.2 at.\% Ag. The primary arm spacing in the dendritic regime is approximately 330 nm, in good agreement with the experimentally measured range of 300–450 nm (Fig. \ref{fig:dendies}). 

To further compare with experiments, we analyzed the Ag concentration profiles across dendrites for 13.9 at.\% Ag. Of particular interest is the ratio of minimum to maximum concentration, which serves as an approximate measure of the partition coefficient at the corresponding solidification velocity. Notably, this ratio for the dendrite morphology exhibiting deep grooves (0.2 m/s at 13.9 at.\% Ag in panel B) corresponds to the equilibrium partition coefficient, since the local interface velocity in the normal direction vanishes at the groove root. As shown in Fig. \ref{fig:microstructures}C,\ref{fig:microstructures}D, with increasing solidification velocity, the minimum-to-maximum concentration ratio rises from approximately 0.4 for dendrites with deep grooves to unity for the planar interface, consistent with the expected behavior of the partition coefficient in the limits of zero and very high velocity. This trend qualitatively agrees with experimental observations (Fig. \ref{fig:dendies}), although the absolute values differ appreciably (0.4–0.8 in the PF simulations versus 0.8–0.9 experimentally).

\begin{figure*}[!h]
\centering
      \captionsetup{singlelinecheck = false, font=footnotesize, labelsep=period}
      \includegraphics[width=0.9\textwidth]{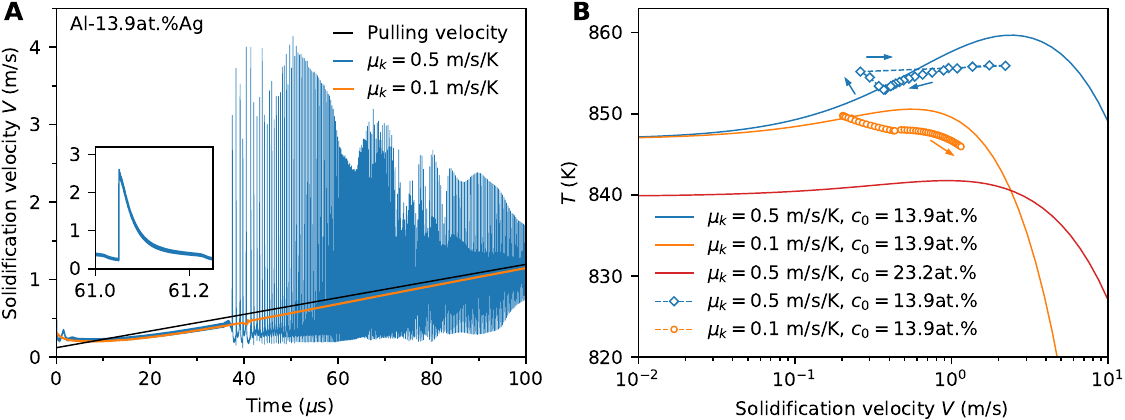}
      \caption{ Dynamics at the solidification front. (A) Pulling velocity and instantaneous interface velocity versus time for 13.9at.\% Ag. The inset shows the solidification velocity within a banding cycle. (B) Interface temperature $T$ plotted against solidification velocity.  Solid lines denote steady-state values; symbols (diamonds and circles) show instantaneous data in PF simulations. The blue diamonds trace a single banding cycle shown in the inset of panel A, and the arrows indicate its temporal progression. }
\label{fig:TV}
\end{figure*}

Fig. \ref{fig:microstructures} further illustrates that reducing $\mu_k^0$ stabilizes the solid-liquid interface, whereas changes in $c_\infty$ have minimal direct impact on interface stability, but significantly affect $V_\text{abs}$. To investigate the stability of the solid-liquid interface in greater detail, we examined the solidification front behavior. Fig. \ref{fig:TV}A shows solidification velocity traces over time, highlighting how instantaneous velocity tracks the pulling velocity closely in dendritic regimes (for both $\mu_k^0=0.1$ and 0.5 m/s/K), whereas velocity oscillations occur in the banded regime. Fig. \ref{fig:TV}B presents the steady-state and instantaneous temperatures at the solidification front as functions of solidification velocity. A lower $\mu_k^0$ notably reduces the temperature peak, effectively suppressing banding cycles that oscillate between high and low velocities, which explains why the banded structures occur at $\mu_k^0=0.5$ m/s/K, but not at 0.1 m/s/K in our simulations (Fig. \ref{fig:microstructures}). Additionally, increasing $c_\infty$ lowers both the peak and baseline temperatures, explaining why higher $c_\infty$ tends to eliminate banded structures (Fig. \ref{fig:microstructures}A,B). The higher ends of the diamonds and orange circles do not lie close to the solid line, because the interface in the PF simulations is non-planar, resulting in lower undercooling compared to a planar interface.


\subsubsection{Sharp interface theory}

While the PF simulations reproduce the experimentally observed decrease in $V_\text{abs}$ with increasing Ag concentration, they do not, by themselves, identify which thermodynamic quantities govern this trend. To develop analytical insight, we revisit the linear stability analysis of the solid-liquid interface for concentrated alloys. A key input to this analysis is the velocity-dependent phase diagram. As shown in Ref. \cite{ji2025phase}, the latter is obtained by seeking 1D steady-state solutions of Eqs. \ref{eq:phi} and \ref{eq:c} in a moving frame at some fixed imposed velocity $V$ and fixed concentration in the liquid far from the interface $c_\infty$, which is equal to the nominal composition. The solution of these equations (Eqs. (50) and (51) in Ref. \cite{ji2025phase}) uniquely fixes the interface growth temperature $T$ and the concentration on the liquid side of the interface $c_l^0(V,T)$; by mass conservation the concentration on the solid side of the interface $c_s^0(V,T)=c_\infty$. Repeating this procedure for different values of $V$ and $c_\infty$ over the entire hypoeutectic range completely determines the functions $c_l^0(V,T)$ and $c_s^0(V,T)$ that characterize the non-equilibrium phase diagram.

\begin{figure*}[!ht]
\centering
      \captionsetup{singlelinecheck = false, font=footnotesize, labelsep=period}
      \includegraphics[width=\textwidth]{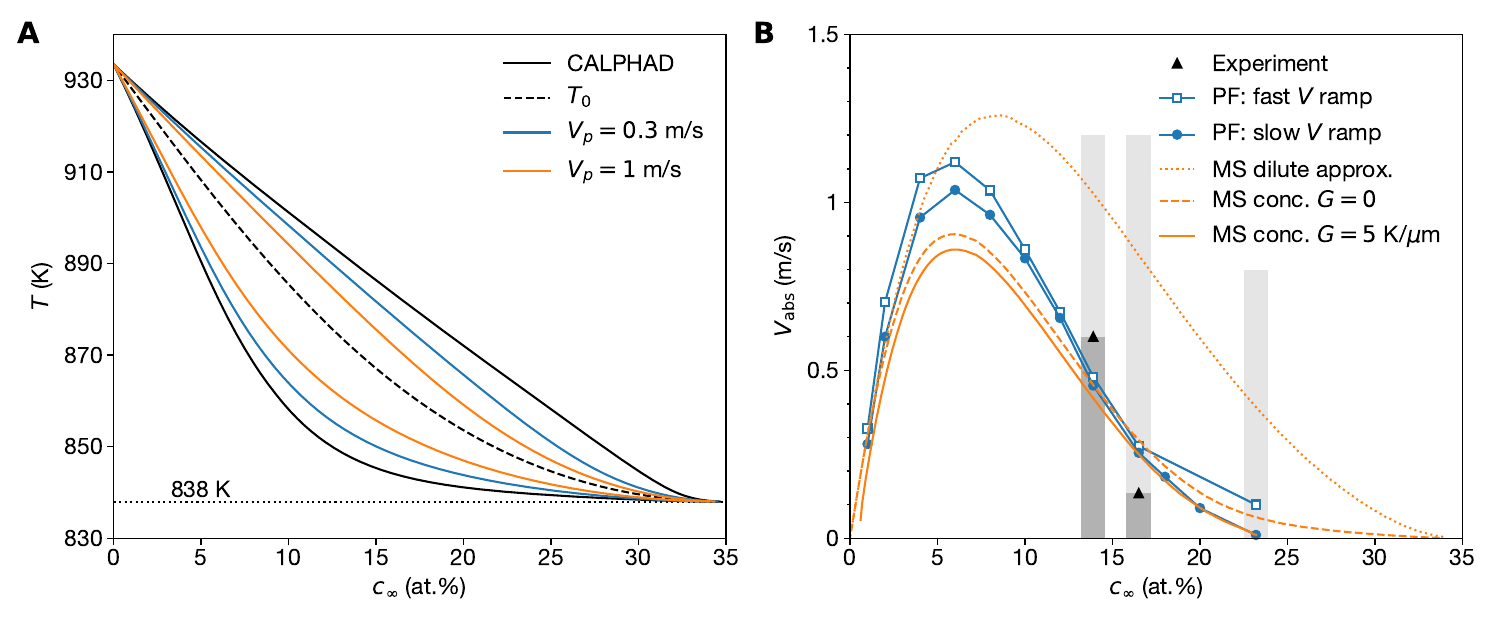}
      \caption{Stability analysis of the solid-liquid interface during rapid solidification. (A) Velocity-dependent phase diagram of hypoeutectic Al-Ag alloys with both solidus and liquidus slopes approaching 0 at the eutectic point. Black solid lines show the equilibrium phase diagram obtained using the CALPHAD method \cite{witusiewicz2004ag}. Blue and orange lines correspond to growth velocities of 0.3 m/s and 1 m/s, respectively, calculated by solving steady-state one-dimensional PF evolution equations without kinetic undercooling \cite{ji2025phase}. The $T_0$ line (dashed) indicates where solid and liquid free-energy densities are equal. (B) Absolute stability velocity ($V_\text{abs}$) as a function of Ag composition. Black triangles represent experimentally estimated $V_\text{abs}$ from thin-film Al-Ag DTEM. Dark and light gray regions denote cellular/dendritic and planar growth regimes, respectively. Blue squares and circles show 2D PF simulation results using the experimentally measured velocity ramp (Fig. \ref{fig:velocity_calcs}C) and a slow velocity ramp, respectively, where the experimental ramp increases from 0.1 m/s. Orange lines represent MS theory predictions under different conditions: dotted line shows standard MS theory under dilute approximation extended to concentrated compositions (Eq. \ref{eq:VabsMullins}); dashed line shows modified MS theory for concentrated alloys with $G=0$; solid line shows modified MS theory with $G=5$ K/$\mu$m. For all results, $\mu_k^0=0.1$ m/(s·K). }
\label{fig:phase_diagram}
\end{figure*}

Fig. \ref{fig:phase_diagram}A presents the resulting phase diagram for hypoeutectic Al-Ag alloys at two representative velocities, shown without kinetic undercooling for clarity. The equilibrium solidus and liquidus lines (solid black lines), derived from CALPHAD calculations~\cite{witusiewicz2004ag}, exhibit increasing and decreasing miscibility gaps at low and high Ag concentrations, respectively. At a velocity of 0.3~m/s (blue lines), the solidus and liquidus lines move closer together, indicating a reduced partition coefficient and the onset of solute trapping. This convergence becomes more pronounced at 1.0~m/s (orange lines), and under sufficiently extreme velocities the phase boundaries are expected to merge toward the $T_0$ line (dashed line), representing entirely suppressed solute segregation. Notably, the narrowing of the miscibility gap at higher concentrations and velocities suggests that $\Delta c^0 = c^0_l - c^0_s$, rather than the nominal composition alone, is the relevant quantity controlling interface stability.

Before deriving a modified stability criterion, we first assess how well the classical MS expression (Eq.~\ref{eq:VabsMullins}) performs when extended to concentrated compositions. Although Eq.~\ref{eq:VabsMullins} is strictly valid only in the dilute limit, we follow recent work by Tourret et al.~\cite{tourret2023morphological} and apply it as a benchmark. Since the liquidus slope $m_l$ and the partition coefficient $k$ in Eq.~\ref{eq:VabsMullins} depend on solidification velocity under rapid solidification, $V_\text{abs}$ must be determined numerically. To this end, we computed the phase diagram for velocities ranging from 0.01 to 1.5~m/s in increments of 0.01~m/s, explicitly accounting for kinetic undercooling. From these results, we determined $m_l = m_l(c_\infty, V)$ and $k = c_\infty / c^0_l(c_\infty, V)$, where $c^0_l$ is the liquid concentration peak at the interface. The calculated $V_\text{abs}$ is shown as the dotted orange curve in Fig.~\ref{fig:phase_diagram}B. At low Ag compositions, $V_\text{abs}$ increases with $c_\infty$ because $V_\text{abs}$ is proportional to $c_\infty$ in the dilute limit (Eq.~\ref{eq:VabsMullins}) while both $m_l$ and $k$ remain nearly constant and only weakly velocity-dependent. At higher solute concentrations, however, $V_\text{abs}$ decreases with increasing composition. Although this trend is qualitatively consistent with experimental measurements (black triangles), a significant quantitative discrepancy remains, indicating that the dilute-alloy assumptions underlying Eq.~\ref{eq:VabsMullins} break down at these compositions.

To address this discrepancy, we derive a modified stability criterion for concentrated alloys by revisiting the free-boundary problem under rapid solidification conditions. In the frame moving at a velocity $V$ along $z$, the concentration profile $c(z)$ satisfies: 
\begin{equation}
\begin{aligned}
D_l\nabla^2 c + V \frac{\partial c}{\partial z} = \frac{\partial c}{\partial t}.
\end{aligned}
\end{equation}
The concentration profile also satisfies the mass conservation condition at the solid-liquid interface,
\begin{equation}
\begin{aligned}
(c_l - c_s)V = -D_l \frac{\partial c}{\partial n}\Big|_l,
\end{aligned}
\end{equation}
where $n$ indicates the direction normal to the interface. From these two equations, one can obtain the steady-state concentration profile in the liquid using $\partial c/\partial t=0$:
\begin{equation}
\begin{aligned}
c(x)=c_s^0+(c_l^0-c_s^0)e^{-\frac{V}{D_l}z}.
\end{aligned}
\end{equation}
Here, $c_l^0$ and $c_s^0=c_\infty$ are the liquid and solid concentrations at the stable interface, which can be obtained from the phase diagram. For a slightly curved solid–liquid interface during rapid solidification, the concentration profile becomes: 
\begin{equation}
\begin{aligned}
c(x) = (c_l^0 - c_s^0)e^{-Vz/D_l} + c_s^0 + \hat{c}_k e^{i\vec{k}\cdot\vec{x} - qz + \omega_k t},
\end{aligned}
\label{eq:perturb}
\end{equation}
where $\hat{c}_k e^{i\vec{k}\cdot\vec{x} - qz + \omega_k t}$ represents a small-amplitude perturbation, $\vec{k}$ is a wave vector perpendicular to $z$, and $\omega_k$ is the growth rate of the amplitude that determines stability. Applying this to the diffusion equation yields:
\begin{equation}
\begin{aligned}
q=\frac{1}{l}+\sqrt{\frac{1}{l^2}+k^2+\frac{\omega_k}{D_l}},
\end{aligned}
\label{eq:q}
\end{equation}
where $l=2D_l/V$ is the diffusion length. Similar to the concentration profile, the interface position can be expressed as:
\begin{equation}
\begin{aligned}
x_\text{int} \equiv \zeta= \hat{\zeta}_k \exp(i\vec{k}\cdot\vec{x} + \omega_k t),
\end{aligned}
\end{equation}
where $\hat{\zeta}_k$ is a small amplitude. 
The concentrations on the solid and liquid sides of the interface are given by:  
\begin{equation}
\begin{aligned}
c_l =& c_l^0 - \frac{\Gamma}{m_l} \kappa - \frac{G}{m_l}\zeta,\\
c_s =& c_s^0 - \frac{\Gamma}{m_s} \kappa- \frac{G}{m_s}\zeta,\\
\end{aligned}
\label{eq:gibbsc}
\end{equation}
where $m_l(V,T)=-1/(\partial c_l^0(V,T)/\partial T)_V$ and $m_s(V,T)=-1/(\partial c_s^0(V,T)/\partial T)_V$ are the magnitudes of the local liquidus and solidus slopes of the non-equilibrium phase diagram, respectively, $\Gamma$ is the Gibbs-Thomson condition assumed constant, and $G$ is the temperature gradient.
In the equations above, the effect of the kinetic undercooling is already intrinsically incorporated into the functions $c_s^0(V,T)$ and $c_l^0(V,T)$, where $T$ is the reference growth temperature of the planar interface determined by the relation $c_s^0(V,T)=c_\infty$. This kinetic contribution cannot be explicitly separated from the concentrations, because $c_s^0(V,T)$ and $c_l^0(V,T)$ are derived from the steady-state solutions of the coupled PF ($\phi$) and diffusion ($c$) equations, where the kinetic dependence is nonlinear \cite{ji2025phase}. Following the standard stability analysis procedure \cite{mullins1964stability, langer1980instabilities}, one can apply Eq. \ref{eq:gibbsc} to the mass conservation condition and linearize the equation using $\kappa=k^2\zeta$ and $V\rightarrow V+\frac{\partial \zeta}{\partial t}=V+\zeta\omega_k$, which yields: 
\begin{equation}
\begin{aligned}
\Delta c^0\omega_k\hat{\zeta}_k-\frac{\Gamma}{\bar{m}}k^2\hat{\zeta}_kV-\frac{G}{\bar m}\hat{\zeta}_kV=-\frac{V^2}{D_l}\Delta c^0 \hat{\zeta}_k+qD_l\hat{c}_k,
\end{aligned}
\label{eq:continuity}
\end{equation}
where $\Delta c^0=c_l^0-c_s^0$ and $1/\bar{m}=1/m_l-1/m_s$. Additionally, one can linearize the non-equilibrium Gibbs-Thomson relation, which gives:
\begin{equation}
\begin{aligned}
-\frac{V}{D_l}\Delta c^0\hat{\zeta}_k+\hat{c}_k=-\frac{\Gamma}{m_l}k^2\hat{\zeta}_k-\frac{G}{m_l}\hat{\zeta}_k.
\end{aligned}
\label{eq:gibbs}
\end{equation}
Solving $\hat{c}_k$ and substituting it into Eq. \ref{eq:continuity} yields the equation:
\begin{equation}
\begin{aligned}
\Delta c^0\omega_k=\frac{\Gamma}{\bar{m}}k^2V+\frac{G}{\bar m}V-\frac{V^2}{D_l}\Delta c^0+ D_lq\left( \frac{V}{D_l}\Delta c^0- \frac{\Gamma}{m_l}k^2 - \frac{G}{m_l} \right).
\end{aligned}
\label{eq:omega}
\end{equation}
As defined in Eq. \ref{eq:perturb}, a positive growth rate ($\omega_k > 0$) indicates perturbation amplification. Consequently, the absolute stability velocity is determined by the stability condition $\omega_k = 0$. We note that, in the present derivation, we have neglected terms on the right-hand-side of Eq. \ref{eq:gibbsc} proportional to $(\partial c_\nu^0(V,T)/\partial V)_Td\zeta/dt$ $(\nu=l,s)$ that do not modify the absolute stability limit because $d\zeta/dt\sim \omega_k\hat \zeta_k$ vanishes exactly at this limit defined by $\omega_k=0$. These extra terms, however, need to be included to describe oscillatory modes underlying banding \cite{karma1993interface} that are observed in the PF simulations at larger kinetic coefficient ($\mu_k=0.5$ m/s/K). Since, here, we only use the present analysis to interpret PF results for lower kinetic coefficient ($\mu_k=0.1$ m/s/K)  where, as in the experiments, banding is absent, we need not include these velocity-dependent corrections in the stability analysis.    

\vspace{12pt}
\noindent\textbf{The isothermal limit}

\noindent In the isothermal limit ($G=0$), Eq. \ref{eq:omega} simplifies considerably. Typically, $k$ is smaller than $1/l$; for example, $1/l = V/2D_l = 1~\text{ms}^{-1}/(2\times 5\times10^{-9}~\text{m}^2\text{s}^{-1}) = 1\times10^8~\text{m}^{-1}$, whereas $k = 2\pi/\lambda \approx 2\pi/300~\text{nm} \approx 2\times10^7~\text{m}^{-1}$. In this regime, $q$ can be expanded to the leading order $q \approx 2/l + k^2 l/2$, so that all $k$-independent terms on the right-hand side of Eq.~\ref{eq:omega} cancel, leaving only contributions of order $k^2$ and higher. The absolute stability velocity is then determined by the condition that the coefficient of the leading $k^2$ term vanishes. This yields a remarkably simple prediction for the absolute stability velocity, which is implicitly defined by the two equations:
\begin{equation}
\begin{aligned}
V_\text{abs}=\frac{D_lm_s(V_\text{abs},T)\Delta c_0(V_\text{abs},T)}{\Gamma},
\end{aligned}
\label{eq:VabsNew}
\end{equation}
and $c_s^0(V_\text{abs},T)=c_\infty$, which together uniquely predict $V_\text{abs}$ and $T$ (the planar front solidification temperature at velocity $V_\text{abs}$), where we have defined the miscibility gap $\Delta c_0=c_l^0(V_\text{abs},T)-c_\infty$.
In the dilute limit, where $m_s=m_l/k$ and $\Delta c^0=c_\infty/k-c_\infty$, the above equation reduces to commonly used Eq. \ref{eq:VabsMullins}, which further validates Eq. \ref{eq:VabsNew}.

\begin{figure*}[!ht]
\centering
      \captionsetup{singlelinecheck = false, font=footnotesize, labelsep=period}
      \includegraphics[width=0.6\textwidth]{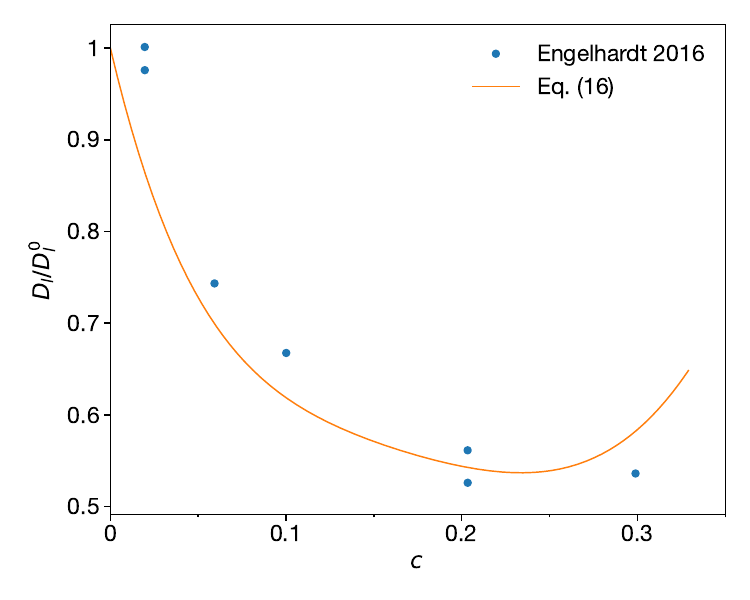}
      \caption{Dependence of liquid diffusivity on Ag composition. Experimental values were measured at 983~K~\cite{engelhardt2016AlAgliqdiffusion}, significantly higher than the $\sim$850~K relevant to this work, which accounts for the difference in absolute diffusivity. To facilitate comparison, the diffusivity ($D_l$) is normalized by its dilute-limit value ($D_l^0$) for both the experimental data and the present model (Eq.~\ref{eq:diffusivity}). }
\label{fig:Dl}
\end{figure*}

Notably, the diffusivity in Eq. \ref{eq:VabsNew} is not constant, but instead depends on concentration. In the present stability analysis, we used a standard diffusion equation for the concentration field. Rewriting the evolution equation from our PF model (Eq. \ref{eq:c}) in the standard form $\partial c/\partial t=\vec{\nabla}\cdot(D_l\vec{\nabla}c)$, we obtain the effective diffusivity: 
\begin{equation}
\begin{aligned}
D_l(c)=D_l^0\frac{c(1-c)}{RT_M}\frac{\partial^2G_l(c,T)}{\partial c^2}.
\end{aligned}
\label{eq:diffusivity}
\end{equation}
In the dilute limit $c \rightarrow 0$, this reduces to $D_l^0 = 5 \times 10^{-9}$~m$^2$/s. As shown in Fig.~\ref{fig:Dl}, the diffusivity calculated using Eq. \ref{eq:diffusivity} decreases with Ag concentration, dropping to approximately $2.66 \times 10^{-9}$~m$^2$/s (i.e., $D_l/D_l^0 = 0.53$) at $\sim$20~at.\% Ag. This concentration-dependent reduction is quantitatively consistent with the experimental measurements of Engelhardt et al. \cite{engelhardt2016AlAgliqdiffusion}, as demonstrated by the good agreement between the normalized diffusivities in Fig. \ref{fig:Dl}. The absolute experimental values are notably larger ($8 \times 10^{-9}$~m$^2$/s in dilute Al-Ag, compared to $5 \times 10^{-9}$~m$^2$/s here), which can be attributed to the higher measurement temperature of 983 K versus $\sim$850 K in the present work. Since $V_\text{abs}$ is directly proportional to $D_l$ (Eq.~\ref{eq:VabsNew}), this roughly twofold decrease in diffusivity at higher Ag concentrations contributes to the reduction of the absolute stability threshold.

Numerical solution of Eq. \ref{eq:VabsNew} using the composition-dependent diffusivity, and velocity-dependent $m_s$ and $\Delta c^0$ from the phase diagram, yields the dashed orange line in Fig. \ref{fig:phase_diagram}B. This new formulation agrees with the classical MS theory in the dilute limit (dashed orange line versus dotted orange line). At higher compositions, this result significantly reduces $V_\text{abs}$ predicted by the standard MS theory to the experimental observation level (black triangles). Additionally, Eq. \ref{eq:VabsNew} reveals the decrease in $V_\text{abs}$ at higher compositions arises from the reduction of miscibility gap $\Delta c^0$, diffusivity $D_l$, and solidus slope $m_l$, where both $\Delta c^0$ and $m_l$ are predominant because $D_l$ decreases by a factor of 2 (Eq. \ref{eq:diffusivity}), while both $m_s$ and $\Delta c^0$ decreases to 0 at the eutectic point.

\vspace{12pt}
\noindent\textbf{Effects of thermal gradients}

\noindent Since $k$ is smaller, but not far smaller, than $1/l$, the most rigorous approach to determine $V_\text{abs}$ at a finite thermal gradient ($G \neq 0$) is to numerically solve Eq. \ref{eq:omega} for the marginal stability condition $\omega_k=0$ subject to Eq. \ref{eq:q}. The calculation yields stability velocities (solid orange line in Fig. \ref{fig:phase_diagram}B) slightly lower than the isothermal limit (dashed orange line). This solution shows excellent agreement with standard MS theory in the dilute regime and matches experimental measurements at higher compositions. A critical feature of this analysis is the existence of a maximum concentration at 23.2 at.\% (where $V_\text{abs} \approx 0.01$ m/s), beyond which no theoretical solution for $V_\text{abs}$ exists. This observation contrasts with the isothermal prediction, where $V_\text{abs}$ decreases gradually to zero as the composition approaches the eutectic point.

\subsubsection{Comparison between theory and PF simulations under velocity ramps}

To compare with the theoretical analysis (Eqs. \ref{eq:omega} and \ref{eq:VabsNew}), we performed PF simulations across a range of nominal compositions to determine the absolute stability velocity. Initially, we imposed the same velocity ramp used in the experiments (Fig. \ref{fig:velocity_calcs}B; also used in Fig. \ref{fig:microstructures}B). The resulting $V_\text{abs}$ values (open squares, Fig. \ref{fig:phase_diagram}B) show quantitative agreement with theory in the dilute limit and qualitative agreement with both theory and experiment at $\sim$15 at.\% Ag. However, significant deviations occur near the peak stability at $\sim$6 at.\% Ag and at the maximum of 23.2 at.\% Ag, where the simulated values exceed theoretical predictions.

We attribute this discrepancy to the rapid velocity change of the experimental ramp (0.1 to 1 m/s within 100 $\mu$s). As shown in Fig. \ref{fig:TV}A (orange vs. black lines), the actual interface solidification velocity lags behind the imposed pulling velocity under these transient conditions. To mitigate this lag and approximate steady-state growth, we conducted additional simulations with a significantly slower ramp (0.1 m/s per 200 $\mu$s—a 18-fold reduction in velocity change rate). These quasi-static simulations (blue circles, Fig. \ref{fig:phase_diagram}B) yield excellent quantitative agreement with the dilute-limit theory. Notably, at high concentrations, the slow-ramp simulations align quantitatively with the theoretical analysis incorporating the thermal gradient, confirming the existence of a maximum composition at approximately 23.2 at.\% Ag. While a deviation of $\sim$20\% persists in the peak $V_\text{abs}$ region, this difference is sufficiently small to establish consistency, thereby validating both the PF model and the theoretical framework.

\subsubsection{Discussion on computational and theoretical analysis}

The morphological stability of the solid-liquid interface during directional solidification has been studied extensively since the foundational analysis of Mullins and Sekerka \cite{mullins1964stability}, with subsequent extensions to rapid solidification by Trivedi and Kurz \cite{TRIVEDI1986planar}. Building on this stability analysis, Kurz, Giovanola, and Trivedi \cite{kurz1986theory} developed the KGT model for dendritic growth kinetics, which has since been extended to multicomponent alloys \cite{rappaz1990analysis, coates1968solid, COLIN2016absoluteternary, tourret2023morphological} and remains one of the most widely used frameworks for predicting microstructure selection across a broad range of solidification conditions. These formulations yield the classical absolute stability limit (Eq. \ref{eq:VabsMullins}) under the assumption that the liquidus slope, partition coefficient, and liquid diffusivity remain approximately constant, a condition valid only for dilute alloys. Our results demonstrate this assumption breaks down for concentrated alloys: the classical theory predicts monotonically increasing $V_{\text{abs}}$ with solute content, whereas our experiments, PF simulations, and modified stability analysis consistently reveal the opposite trend in hypoeutectic Al-Ag alloys.

The modified stability criterion derived here (Eq. \ref{eq:VabsNew} and \ref{eq:omega}) identifies the miscibility gap $\Delta c^0 = c_l^0 - c_s^0$ as the fundamental thermodynamic quantity governing interface instability. In the dilute limit, where $m_s = m_l/k$ and $\Delta c^0 = c_\infty(1-k)/k$, Eq. \ref{eq:VabsNew} reduces exactly to the classical Mullins-Sekerka expression, providing internal validation. At higher concentrations, however, the two formulations diverge because the miscibility gap narrows, the solidus slope changes, and the diffusivity decreases — effects absent from dilute-limit theories. Crucially, the values of $\Delta c^0$, $m_s$, and $D_l(c)$ entering Eq. \ref{eq:VabsNew} are computed directly from one-dimensional, steady-state phase-field solutions using CALPHAD free energies \cite{witusiewicz2004ag, dinsdale1991sgte}, rather than approximated via closed-form trapping laws such as the continuous growth model \cite{aziz1982model, aziz1988continuous}. Using these inputs, the sharp-interface theory quantitatively reproduces the experimentally measured $V_{\text{abs}}$ values across the entire composition range studied (Fig. \ref{fig:phase_diagram}B). The theoretical predictions also align with our 2D PF simulations performed under a slow velocity ramp approximating steady-state conditions, establishing three-way agreement between experiment, theory, and simulation.

Recent work by Tourret et al. \cite{tourret2023morphological} on Ni-Mo-Al alloys similarly identified shortcomings of the KGT model, attributing the discrepancy primarily to kinetic undercooling, which raises the effective $V_{\text{abs}}$ above classical predictions. Our findings are complementary: in the Al-Ag system, it is the concentrated-alloy thermodynamics, specifically the vanishing miscibility gap near the eutectic, that dominates the departure from classical predictions. Both studies converge on the necessity of properly accounting for the velocity-dependent phase diagram, but the controlling mechanisms differ depending on the alloy system and composition regime.

The role of the kinetic coefficient $\mu_k^0$ connects to the banding instability framework of Karma and Sarkissian \cite{karma1993interface}, who showed that relaxation oscillations driven by solute trapping produce banded structures near absolute stability. Our simulations show reducing $\mu_k^0$ from 0.5 to 0.1 m/s/K suppresses banding by lowering the peak of the steady-state $T(V)$ curve (Fig. \ref{fig:TV}), providing a new quantitative insight: the kinetic coefficient acts as a control parameter for banding suppression, complementing the roles of composition and latent-heat diffusion identified previously \cite{kurz1993AlCu, kurz1995AlCu, karma1993interface}.

However, the requirement of $\mu_k^0 \approx 0.1$ m/s/K, which is substantially below typical literature values of 0.3--1 m/s/K for metals \cite{mendelev2010molecular, monk2009determination, hoyt2002atomistic}, warrants further experimental or atomistic confirmation. A significant clue emerges from comparing experimental and simulated solute partitioning: the measured ratio of minimum to maximum Ag concentration across dendrites (0.8--0.9) is substantially higher than in the PF predictions (0.4--0.8) at comparable velocities (Figs. \ref{fig:dendies} and \ref{fig:microstructures}D). This observation indicates that the effective partition coefficient is closer to unity in experiments than in the model, implying a smaller miscibility gap at high velocities than predicted by the CALPHAD-derived free energies.

Two factors likely contribute to this discrepancy. The first is the limited accuracy of the CALPHAD-based thermodynamic description in the far-from-equilibrium regime. Since CALPHAD assessments are constrained primarily by equilibrium data \cite{witusiewicz2004ag}, their extrapolation to the high-velocity regime near the $T_0$ line may overestimate the miscibility gap. A smaller miscibility gap would reduce the driving force for both morphological instability and banding, stabilizing the planar interface without requiring an anomalously low $\mu_k^0$. The need for a reduced kinetic coefficient in the present simulations thus likely compensates, at least in part, for an overestimated miscibility gap at high velocities. More reliable thermodynamic descriptions in this regime, informed by atomistic simulations or dedicated rapid solidification experiments, would improve quantitative predictions of microstructure selection under AM conditions.

The second factor is the dimensionality effect of the present simulations, which do not capture the three-dimensional (3D) geometry of the thin-film samples (thickness $\sim$ 100 nm). In the experiments, the interaction between the solidified alloy and the substrate gives rise to a contact angle at the substrate surface, which acts to stabilize the solid-liquid interface. Moreover, in this 3D geometry, solute-enriched liquid can bypass the advancing solidification front by flowing along the fully wetted top surface of the sample or through liquid channels that do not sever the solidified alloy from the substrate. This contrasts with the 2D case, where deep interdendritic grooves must form at low velocities to allow solute-enriched liquid to pass. Consequently, the Ag concentration averaged over the film thickness in the grooved regime is expected to be substantially lower than in the 2D simulations, yielding a smaller apparent partition coefficient that is more consistent with the experimental measurements (Fig. \ref{fig:microstructures}D). Further studies involving 3D PF simulations that explicitly incorporate the thin-film geometry, substrate wetting boundary conditions, and contact angle effects, together with experimental measurements of Ag concentration across thickness using atom probe tomography or depth-resolved energy dispersive spectroscopy, would be valuable for quantitatively testing this hypothesis.


\section{Conclusions}


Experiments on concentrated Al–Ag alloys in thin films reveal a decreasing transition velocity ($V_{\mathrm{abs}}$) from cellular/dendritic to partitionless planar structures with increasing Ag composition. As $V_{\mathrm{abs}}$ decreases, the interface becomes increasingly stable, such that lower (13.9 at\% Ag) compositions experience the dendritic/cellular to planar transition at a measured velocity of 0.6 $\pm$ 0.1 m/s, while the higher compositions (16.5, 23.2 at\% Ag) have a sufficiently low transition velocity that these experiments did not capture it. In effect, solidification occurs entirely as a partitionless planar front. There is still an observable increase in interface stability from 16.5 to 23.2 at\% Ag; remelting produces a destabilizing effect that produces cellular growth at 16.5 at\% Ag but not 23.2 at\% Ag, indicating the higher composition produces a more stable interface. 

We extended the linear stability analysis of the planar interface to concentrated alloys by using two ingredients: a velocity-dependent. non-equilibrium phase diagram computed from 1D moving-front solutions of the PF evolution equations, and a concentration-dependent liquid diffusivity. The analysis predicts a non-monotonous behavior of $V_{\text{abs}}$ as a function of alloy concentration, which is characterized by an increase at small concentration followed by a decrease at larger concentration. At a qualitative level, this behavior reflects the non-monotonous behavior of the miscibility gap (or freezing range) as a function of concentration. This effect is particularly pronounced in the Al-Ag system, as the solidus has an inflection point where it drastically decreases in slope, which also deflects the $T_0$ line slope. At a quantitative level, the predictions of linear stability analysis are in good quantitative agreement with both PF simulations over the entire hypoeutectic range of concentration, and with experimental observations for the three concentrated alloys that were investigated here. These results improve the theoretical basis for predicting microstructure selection in general in binary alloys processed via AM. The extension of the present approach to multicomponent alloys is an interesting future prospect.

\section*{CRediT Authorship Contribution Statement} 

B.R.: Conceptualization, Formal Analysis, Investigation, Methodology, Validation, Visualization, Writing - Original Draft, Writing - Review \& Editing. M.Z.: Conceptualization, Formal Analysis, Investigation, Methodology, Validation, Visualization, Writing - Original Draft, Writing - Review \& Editing. T.L.:  Writing - Original Draft, Writing - Review \& Editing. J.R.: Formal Analysis, Investigation, Methodology, Validation, Writing - Review \& Editing. J.T.M: Conceptualization, Formal Analysis, Investigation, Methodology, Validation, Writing - Review \& Editing. A.K.: Conceptualization, Formal Analysis, Funding acquisition, Investigation, Methodology, Project administration, Resources, Supervision, Validation, Visualization, Writing - Review \& Editing. A.J.C.: Conceptualization, Formal Analysis, Funding acquisition, Investigation, Methodology, Project administration, Resources, Supervision, Validation, Visualization, Writing - Review \& Editing. 

\section*{Declaration of Competing Interests} 

The authors declare no competing interests.

\section*{Acknowledgments}

This work was funded by the U.S Department of Energy, Office of Science, Office of Basic Energy Sciences under Award Numbers DE-SC0020870 (A.J.C.) and DE-SC0020895 (A.K.). B.R was also partially supported by the National Nuclear Security Administration Laboratory Residency Graduate Fellowship. A.J.C. acknowledges support by the U.S. Department of Energy through the Los Alamos National Laboratory during the preparation of this manuscript. Los Alamos National Laboratory is operated by Triad National Security, LLC, for the National Nuclear Security Administration of the U.S. Department of Energy (Contract No. 89233218CNA000001). J.T.M. and J.R. performed work under the auspices of the U.S. Department of Energy by Lawrence Livermore National Laboratory under contract No. DE-AC52-07NA27344.

\bibliographystyle{naturemag}
\bibliography{reference}

\end{document}